\documentclass[pra]{revtex4-2}

\usepackage{graphicx}
\usepackage{color}

\begin{document}



\title{Characterization of the non-classical relation between measurement outcomes represented by non-orthogonal quantum states}

\author{Ming Ji}
\author{Holger F. Hofmann}
\email{hofmann@hiroshima-u.ac.jp}
\affiliation{
Graduate School of Advanced Science and Engineering, Hiroshima University,
Kagamiyama 1-3-1, Higashi Hiroshima 739-8530, Japan
}

\begin{abstract}
Quantum mechanics describes seemingly paradoxical relations between the outcomes of measurements that cannot be performed jointly. In Hilbert space, the outcomes of such incompatible measurements are represented by non-orthogonal states. In this paper, we investigate how the relation between outcomes represented by non-orthogonal quantum states differs from the relations suggested by a joint assignment of measurement outcomes that do not depend on the actual measurement context. The analysis is based on a well-known scenario where three statements about the impossibilities of certain outcomes would seem to make a specific fourth outcome impossible as well, yet quantum theory allows the observation of that outcome with a non-vanishing probability. We show that the Hilbert space formalism modifies the relation between the four measurement outcomes by defining a lower bound of the fourth probability that increases as the total probability of the first three outcomes drops to zero. Quantum theory thus makes the violation of non-contextual consistency between the measurement outcomes not only possible, but actually requires it as a necessary consequence of the Hilbert space inner products that describe the contextual relation between the outcomes of different measurements.
\end{abstract}

\maketitle
\section{Introduction}
As shown by the Bell-Kochen-Specker theorem, quantum theory is not consistent with a measurement independent assignment of outcomes to a set of different measurements that cannot be performed jointly \cite{Bel64,Koc68}. This inconsistency between non-contextual models and quantum theory is usually characterized as a violation of an inequality based on the statistics of a set of incompatible measurements \cite{Cab08,Bad09, Kle12,Pan13,Ysu15,Kun15,Pxu16, Kri17,Kun18,Sch18,Lei20}. The best known of these is the violation of Bell's inequalities, which is widely regarded as convincing evidence that quantum statistics cannot be explained by local hidden variable theories that assign outcomes to each measurement independent of whether the measurement is performed or not \cite{Bel66,Gen05,Zel07,Car18,Tem19}. It is therefore sufficiently clear that quantum superpositions do not represent classical alternatives. However, it is still somewhat unclear how the relations between different measurements are modified by the Hilbert space formalism, since each measurement probability is determined separately and the role that quantum coherence plays in defining the relation between different measurement contexts is represented by abstract algebraic relations that have no obvious analog in Boolean logic.

In the present paper, we investigate the relation between the outcomes of incompatible measurements described by the Hilbert space formalism and show that the non-orthogonality of the state vectors representing the measurement outcomes necessarily results in a violation of non-contextual logic. The present analysis is based on the demonstration of inconsistencies between different measurements presented by Frauchiger and Renner \cite{Fra18}. This consistency paradox seems to represent a fundamental structure of quantum theory since it is mathematically equivalent to both Hardy's paradox \cite{Har92,Har93,Cab13,Mar14} and a Bell's inequality violation \cite{Mer94,Pro19}. The formulation of Frauchiger and Renner is particularly convenient because it is based on statements that correspond to probabilities of zero for specific measurement outcomes. In non-contextual logic, such statements can be used to eliminate combinatoric possibilities. In the case of the Frauchiger-Renner scenario, the joint validity of three statements appears to require the validity of a fourth statement relating to a probability of zero for a fourth measurement outcome. However, the Hilbert space formalism represents these statements as orthogonality relations. The first three orthogonality relations can be satisfied by only one state in the four-dimensional Hilbert space of the quantum system, and this state necessarily violates the fourth statement with a non-vanishing probability of exactly 1/12. We therefore find that the violation of non-contextual logic is a direct consequence of the non-orthogonality of the Hilbert space vectors representing the outcomes of different measurements. Specifically, the representation of measurement outcomes by Hilbert space vectors replaces the combinatoric arguments of non-contextual logic with the more precise relations between vectors defined by their inner products. These inner products determine a necessary violation of non-contextual logic that can be expressed in terms of quantitative bounds for the probabilities corresponding to the different statements in the consistency paradox.

In the following, we analyze the Hilbert space relation between the four statements that define the consistency paradox and derive a sufficiently tight bound for the probability that expresses the violation of non-contextual logic. Specifically, we replace the probabilities of zero corresponding to precise statements with a probability representing the errors in the three initial statements. Since the three statements cannot be verified jointly, this probability is a sum of three measurement probabilities represented by non-orthogonal projection operators. Our analysis thus demonstrates the counter intuitive nature of the Hilbert space relation between measurement outcomes represented by non-orthogonal states. The minimal probability of violating non-contextuality by obtaining the fourth measurement result is then found by analyzing the optimal quantum coherences between the eigenstates of the operator representing the probability sum. The method used to derive the bound thus illustrates the extent to which quantum interference effects determine the relation between the measurement probabilities obtained in different measurement contexts. The analysis presented below shows that quantum contextuality is closely related to the relation between non-orthogonal state vectors in Hilbert space, resulting in seemingly paradoxical results because the probability of one measurement outcome depends on the possible coherences between the eigenstates representing another set of measurement outcomes. We therefore believe that the results presented in this paper can greatly improve our understanding of the paradoxical relation between the statistics of incompatible measurements.

The rest of the paper is organized as follows. In Sec. \ref{sec:paradox}, we formulate the consistency paradox in terms of three statements that can be represented by probabilities of zero and show that the corresponding orthogonality relations necessarily violate non-contextuality. In Sec. \ref{sec:Hilbertspace}, we derive the quantitative relation between the measurement probabilities that make a violation of non-contextuality necessary. In Sec. \ref{sec:statistcalbounds}, we formulate quantitative lower bounds for the probabilities that describe the violation of non-contextuality. It is shown that the probability of violating the fourth statement can only be reduced by increasing the probability of violating one of the three initial statements, highlighting the difference between the relation of statements in Hilbert space and non-contextual logic. Section \ref{sec:conclusion} summarizes the results and concludes the paper.

\section{The consistency paradox}
\label{sec:paradox}

Although the violation of an inequality provides compelling quantitative evidence for the validity of the Bell-Kochen-Specker theorem, the logical structure of the underlying paradox is always represented by a contradiction between a set of precise statements. It is therefore most instructive to consider paradoxical situations where the statements can be identified directly with experimentally accessible measurement outcomes. A contradiction can then be certified whenever a state that always satisfies a set of initial conditions results in a measurement outcome that seems to contradict the statements associated with these initial conditions. A particularly clear description of such a contradiction was provided by Frauchiger and Renner \cite{Fra18}, who discuss the internal consistency of quantum mechanics based on the idea that superpositions represent alternative possibilities even when there is no external measurement. Here, we focus on the structure of the argument, which the scenario shares with Hardy's paradox \cite{Har92,Har93}. Since we would like to focus on the relation between the different possible measurement outcomes defined by their Hilbert space formalism, we refer to this scenario as the consistency paradox. Although the Hilbert space formalism provides a consistent description of all possible measurements, it is not possible to reconcile the measurement outcomes of different measurement contexts with each because the Hilbert space formalism does not allow the simultaneous observation of outcomes described by Hilbert space vectors that are neither orthogonal nor parallel to each other in Hilbert space, preventing any physically meaningful assignment of joint realities to such outcomes.

The consistency paradox concerns a pair of identical physical systems 1, 2 with two observable properties ${F}_{i}$ and ${W}_{i}$ ($i=1,2$) each, where the property ${F}_{i}$ is either 0 or 1 and the property ${W}_{i}$ is either $a$ or $b$. The goal is to identify the relations between
${F}_{i}$ and ${W}_{i}$ without making any direct statements about the relation between ${F}_{i}$ and ${W}_{i}$ in the same system. This is achieved by considering only statements about the correlations between the two systems. Suppose the following three statements are true:
\begin{enumerate}
\item If ${W}_{1}=a$, then ${F}_{2}=1$; ($P_{\mathrm {WF}}(a,0)=0$)
\item If ${W}_{2}=a$, then ${F}_{1}=1$; ($P_{\mathrm {FW}}(0,a)=0$)
\item ${F}_{1}$ and ${F}_{2}$ cannot both be 1, ($P_{\mathrm {FF}}(1,1)=0$)
\end{enumerate}
where the probabilities of the form $P_{\mathrm {mn}}(x,y)$ in the statements (1)-(3) are the joint probabilities of the possible measurement outcomes $x$ of property $m$ in system 1 and $y$ of property $n$ in system 2. We can then use the following logic to conclude that the two properties $W_{1}$ and $W_{2}$ cannot both be $a$ at the same time ($P_{\mathrm {WW}}(a,a)=0$). According to statements (1) and (2), $W_{1}$ and $W_{2}$ can only be $a$ when $F_{1}$ and $F_{2}$ are both $1$. However, this conclusion contradicts statement (3). Therefore, any observation of the outcome $(a,a)$ contradicts at least one of the three statements above. It is possible to express this precise relation between the statements in terms of the statistics of the outcomes $(a,a)$, $(a,0)$, $(0,a)$ and $(1,1)$. Since the outcome $(a,a)$ can only be obtained when one of the statements (1), (2) and (3) is false, the probability of finding $(a,a)$ should not be larger than the probability of finding an error in the statements (1), (2) and (3). Using the probabilities of the outcomes $(a,0)$, $(0,a)$ and $(1,1)$, the probability of $(a,a)$ should satisfy the inequality
\begin{equation}
\label{eq:paa1}
P_{\mathrm {WW}}(a,a)\leq P_{\mathrm {WF}}(a,0)+P_{\mathrm {FW}}(0,a)+P_{\mathrm {FF}}(1,1).
\end{equation}
This inequality defines the bound that applies if the outcomes $(a,a)$, $(a,0)$, $(0,a)$ and $(1,1)$ are related to each other by non-contextual logic. Conversely, a violation of this inequality indicates that it is not possible to assign outcomes to $F_{i}$ when $W_{i}$ is obtained instead and vice versa, even if the correlations between the outcomes in the two systems seems to permit it. It may be worth noting that the formulation of non-contextual logic given above implicitly refers to the quantum mechanics of the situation by ruling out a joint measurement of $F_{i}$ and $W_{i}$ in the same system $i$. Each system is therefore represented by a two dimensional Hilbert space, where the measurement outcomes of each measurement represents a complete basis. The reason for the failure of non-contextual logic in quantum mechanics can be found in the very specific relations between the eigenstates of the non-commuting operators $\hat{F}_{i}$ and $\hat{W}_{i}$ that represent the physical properties in the quantum formalism.

As shown by Frauchiger and Renner, quantum systems can satisfy all three statements (1), (2) and (3) with precision, so that the right hand side of the inequality (\ref{eq:paa1}) is zero. However, the probability of finding the outcome $(a,a)$ will not be zero, resulting in a particularly striking violation of the inequality. We will now consider the quantum mechanics of the scenario by representing the two systems $1,2$ as two-level systems where the outcomes $|0\rangle$ and $|1\rangle$ are the eigenstates of the operators $\hat{F}_{i}$, and $|a\rangle$ and $|b\rangle$ are the eigenstates of the operators $\hat{W}_{i}$. Since the eigenstates are defined in the same two dimensional Hilbert space, they must be related to each other by superpositions. The closest analog to the assumption that the outcomes can be defined independently is the assumption that the eigenstates are mutually unbiased, so the relation between them can be defined as
\begin{eqnarray}
\label{eq:ab}
|a\rangle&=&\frac{1}{\sqrt{2}}\left(|0\rangle-|1\rangle\right)
\nonumber\\
|b\rangle&=&\frac{1}{\sqrt{2}}\left(|0\rangle+|1\rangle\right).
\end{eqnarray}
The eigenstates of the two operators overlap in Hilbert space, describing a specific relation between them that is fundamentally different from the relations of non-contextual logic, where it is implied that different combinations of $(a,b)$ and $(0,1)$ can describe  a non-contextual joint reality of $\hat{F}_{i}$ and $\hat{W}_{i}$. However, a direct observation of the relation between $\hat{F}_{i}$ and $\hat{W}_{i}$ is impossible because of the incompatibility of their measurements. This impossibility is particularly visible in the relation between the outcomes $(a,0)$ and $(0,a)$ in statements (1) and (2). These outcomes are represented by non-orthogonal states with an overlap of $\langle a,0|0,a\rangle=1/2$ in the four dimensional Hilbert space of the two systems. This overlap indicates that the statements cannot represent a joint reality, since it is impossible to obtain the outcomes $(a,0)$ and $(0,a)$ in the same measurement.

We can explore the consequences of this difference between non-contextual logic and the Hilbert space formalism by formulating the statements (1), (2) and (3) in terms of their Hilbert space relations. Since all three statements refer to probabilities of zero for specific outcomes, it is possible to represent them as orthogonality relations between specific directions in the four dimensional Hilbert space and the initial state $|\phi_{0}\rangle$,
\begin{enumerate}
\item $\langle a,0|\phi_{0}\rangle=0$ $(P_{\mathrm {WF}}(a,0)=0)$, therefore $\langle 0,0|\phi_{0}\rangle=\langle 1,0|\phi_{0}\rangle$;
\item $\langle 0,a|\phi_{0}\rangle=0$ $(P_{\mathrm {FW}}(0,a)=0)$, therefore $\langle 0,0|\phi_{0}\rangle=\langle 0,1|\phi_{0}\rangle$;
\item $\langle 1,1|\phi_{0}\rangle=0$ $(P_{\mathrm {FF}}(1,1)=0)$.
\end{enumerate}
Since the two systems are described by a four dimensional Hilbert space, the three conditions uniquely define the state $|\phi_{0}\rangle$ as the only state that is orthogonal to all three measurement outcomes. In the ($\hat{F}_1, \hat{F}_2$)-basis, the state is given by
\begin{equation}
\label{eq:phi}
|\phi_{0}\rangle=\frac{1}{\sqrt{3}}(|0,0\rangle+|0,1\rangle+|1,0\rangle).
\end{equation}
Similar to non-contextual logic, the Hilbert space relations corresponding to statements (1), (2) and (3) also give a precise and well-defined solution. However, this solution results in a finite probability for the outcome $(a,a)$ defined by the inner product of the Hilbert space vectors $|a,a\rangle$ and $|\phi_{0}\rangle$. The result is not zero, but
\begin{equation}
\label{eq:paa2}
P_{\mathrm {WW}}(a,a)=\frac{1}{12}.
\end{equation}
The non-orthogonal relation between the states $|a,0\rangle$ and $|0,a\rangle$ thus results in a specific probability of finding $(a,a)$. This result apparently contradicts the non-contextual logic according to which $(a,a)$ can only be obtained when one of the three statements (1)-(3) is false. In quantum mechanics, the orthogonality conditions corresponding to the three statements actually require a non-vanishing probability of $(a,a)$ by defining a specific quantum state that is {\it not} orthogonal to $|a,a\rangle$. The orthogonality relations given above are therefore fundamentally different from non-contextual statements of impossibility, despite the apparent similarity in their experimental observation. Even in the limit of precise statements, quantum mechanics cannot be interpreted by identifying the measurement outcomes with measurement independent realities. For the Hilbert space formalism, this means that we cannot identify the components of a state vector with physical properties of the system unless these physical properties are actually measured. The focus of our investigation should therefore be the role of quantum superpositions in the definition of the relation between the outcomes of different measurements.

The practical problem of identifying the actual relation between different measurements is that Hilbert space inner products can only be observed as measurement probabilities. Unless the probabilities are 0 or 1, a quantitative statistical approach is needed to make sense of the relation. It is therefore useful to consider the type of quantitative relation that the Hilbert space formalism establishes between the probability $P_{\mathrm {WW}}(a,a)$ and the three probabilities $P_{\mathrm {WF}}(a,0)$, $P_{\mathrm {FW}}(0,a)$ and $P_{\mathrm {FF}}(1,1)$. We have already seen that a probability of precisely $P_{\mathrm {WW}}(a,a)=1/12$ is obtained when the three probabilities are zero. We can now formulate a corresponding relation that also applies when the probabilities of the outcomes $(a,0)$, $(0,a)$ and $(1,1)$ are non-zero.

\section{Hilbert space relations between incompatible measurement outcomes}
\label{sec:Hilbertspace}
In the previous section we have seen that the inner products of Hilbert space expression modify the relations between the statements associated with the corresponding measurement outcomes. The reason why these inner products are non-zero is that the statements belong to incompatible measurement contexts and cannot be confirmed jointly in a single measurement of both $\hat{W}$ and $\hat{F}$. As we have seen in the previous section, the quantum formalism requires a non-vanishing value of $P_{\mathrm {WW}}(a,a)=1/12$ for the validity of the three statements (1)-(3). It is therefore necessary to violate at least one of these conditions in order to suppress the probability $P_{\mathrm {WW}}(a,a)$ to zero. Different from non-contextual logic, the Hilbert space formalism thus defines the relation between measurement outcomes as a quantitative relation between their probabilities.
In the following, we will investigate the quantitative relation between the probabilities $P_{\mathrm {WF}}(a,0)$, $P_{\mathrm {FW}}(0,a)$ and $P_{\mathrm {FF}}(1,1)$ corresponding to the statements (1)-(3) and the probability $P_{\mathrm {WW}}(a,a)$ described by the Hilbert space formalism. In particular we will be interested in the minimal value of $P_{\mathrm {WW}}(a,a)$ required in the limit of low values of the probability sum,
\begin{equation}
\label{eq:psum}
P_{\mathrm {S}}=P_{\mathrm {WF}}(a,0)+P_{\mathrm {FW}}(0,a)+P_{\mathrm {FF}}(1,1).
\end{equation}
As the indices show, the three probabilities in this sum refer to measurement outcomes obtained in three different measurement contexts. Specifically, $(a,0)$ and $(0,a)$ could both be true at the same time, since they do not exclude each other. This means that the probability sum could have a value of up to two within the bounds of non-contextual logic.

The probability sum in Eq.(\ref{eq:psum}) corresponds to a collective measure of the violation of statements (1)-(3), with the $P_{\mathrm {S}}$ being the minimal probability that all three statements are correct. The inequality given in Eq. (\ref{eq:paa1}) then simplifies to $P_{\mathrm {WW}}(a,a)\leq P_{\mathrm {S}}$ and the corresponding relations in Hilbert space can be investigated using the quantum mechanical expressions for these two quantities. Specifically, the probability sum $P_{\mathrm {S}}$ can be written as the expectation value of a single operator $\hat{\Pi}_{\mathrm {S}}$ defined by the sum of the three projection operators corresponding to the outcomes $(a,0)$, $(0,a)$ and $(1,1)$,
\begin{eqnarray}
\label{eq:pi}
\hat{\Pi}_{\mathrm S}=|a,0\rangle\langle a,0|+|0,a\rangle\langle 0,a|+|1,1\rangle\langle 1,1|.
\end{eqnarray}
It is worth noting that this operator summarizes projectors that do not commute. Specifically, the non-orthogonality of the states $|a,0\rangle$ and $|0,a\rangle$ means that it is not possible to assign outcomes of zero or one to both projectors at the same time. However, the Hilbert space formalism still assigns eigenvalues and eigenstates to the sum of all three projectors. Of particular interest is the state $|\phi_{0}\rangle$ defined by a probability sum of $P_{\mathrm S}=0$. Since this is an extremal value of the probability sum, it represents an eigenvalue of $\hat{\Pi}_{\mathrm S}$ with $|\phi_{0}\rangle$ as its eigenstate,
\begin{equation}
\label{eq:pi1}
\hat{\Pi}_{\mathrm S}|\phi_{0}\rangle=0.
\end{equation}
This eigenstate is non-degenerate, so it is uniquely defined by the probability sum of $P_{\mathrm S}=0$. We can now find the remaining eigenstates of the operator $\hat{\Pi}_{\mathrm S}$ and express it in its spectral decomposition,
\begin{equation}
\label{eq:pi2}
\hat{\Pi}_{\mathrm S}=\frac{1}{2}|\nu_1\rangle\langle\nu_1|+|\nu_2\rangle\langle\nu_2|+\frac{3}{2}|\nu_3\rangle\langle\nu_3|.
\end{equation}
As Eq. (\ref{eq:pi1}) and the spectral decomposition show, the probability sum has eigenvalues of 0, 1/2, 1 and 3/2. The corresponding eigenstates are given by $|\phi_{0}\rangle$ shown in Eq. (\ref{eq:phi}) and
\begin{eqnarray}
\label{eq:nu}
|\nu_{1}\rangle&=&\frac{1}{\sqrt{2}}\left(|0,1\rangle-|1,0\rangle\right),
\nonumber \\
|\nu_{2}\rangle&=&|1,1\rangle,
\nonumber \\
|\nu_{3}\rangle&=&\frac{1}{\sqrt{6}}\left(2|0,0\rangle-|0,1\rangle-|1,0\rangle\right).
\end{eqnarray}
Note that the state $|1,1\rangle$ is an eigenstate of the probability sum because it is orthogonal to the states $|a,0\rangle$ and $|0,a\rangle$ representing the other two outcomes.
Neither $|a,0\rangle$ nor $|0,a\rangle$ are eigenstates of the operator $\hat{\Pi}_{\mathrm S}$. This is a direct consequence of the overlap between the two non-orthogonal states. The sum of the two projectors $|a,0\rangle\langle a,0|$ and $|0,a\rangle\langle 0,a|$ results in eigenstates that are linear combinations of the two states $|a,0\rangle$ and $|0,a\rangle$, where $|\nu_{1}\rangle$ represents destructive interferences between the two states resulting in an eigenvalue of $1/2$ and $|\nu_{3}\rangle$ represents constructive interference resulting in an eigenvalue of $3/2$. The precision of these eigenvalues would seem to suggest that it is somehow possible that exactly $1/2$ or exactly $3/2$ of the two outcomes $(a,0)$ and $(0,a)$ are true, defying the conventional limitation of truth values to $0$ or $1$. The eigenvalues of the probability sum operator $\hat{\Pi}_{\mathrm S}$ are therefore a clear indication of the contextual nature of the statements involved. Since the two outcomes $(a,0)$ and $(0,a)$ never appear in the same measurement context, the sum of their truth value operators given by $\hat{\Pi}_{\mathrm S}$ can only be observed in a third measurement context that contains neither $|a,0\rangle$ nor $|0,a\rangle$. Since we are only interested in the probability sum and not in the individual contributions to it, this third context is more suitable to describe the relation between $P_{\mathrm S}$ and $P_{\mathrm {WW}}(a,a)$ than the individual statements regarding the outcomes $(a,0)$ and $(0,a)$.

The correct relation between $P_{\mathrm S}$ and $P_{\mathrm {WW}}(a,a)$ can be derived by expressing a possible quantum state $|\psi\rangle$ as a superposition of the eigenstates of the operator $\hat{\Pi}_{\mathrm S}$,
\begin{equation}
\label{eq:psi}
|\psi\rangle=C_{0}|\phi_{0}\rangle+C_{1}|\nu_{1}\rangle+C_{2}|\nu_{2}\rangle+C_{3}|\nu_{3}\rangle,
\end{equation}
where the coefficients $C_{0}$, $C_{1}$, $C_{2}$ and $C_{3}$ are the probability amplitudes of the four eigenstates. The probability sum $P_{\mathrm S}$ can then be expressed using the eigenvalues of $\hat{\Pi}_{\mathrm S}$ and the probabilities associated with these coefficients,
\begin{eqnarray}
\label{eq:PS1}
P_{\mathrm S}=\frac{1}{2}|C_{1}|^{2}+|C_{2}|^{2}+\frac{3}{2}|C_{3}|^{2}.
\end{eqnarray}
Since the eigenstates of the operator $\hat{\Pi}_{\mathrm S}$ form a complete orthogonal basis of the four dimensional Hilbert space, we can also express the state $|a,a\rangle$ representing the measurement outcome of a measurement of $\hat{W}_{1}$ and $\hat{W}_{2}$ in this basis,
\begin{eqnarray}
\label{eq:aa}
|a,a\rangle=-\frac{1}{2\sqrt{3}}|\phi_{0}\rangle+\frac{1}{2}|\nu_{2}\rangle+\sqrt{\frac{2}{3}}|\nu_{3}\rangle.
\end{eqnarray}
Note that $|a,a\rangle$ is orthogonal to $|\nu_{1}\rangle$ because of the symmetry between system 1 and system 2. It is therefore possible to express the probability $P_{\mathrm {WW}}(a,a)$ using the three probability amplitudes $C_{0}$, $C_{2}$ and $C_{3}$,
\begin{eqnarray}
\label{eq:paa3}
P_{\mathrm {WW}}(a,a)=\left|\frac{1}{2\sqrt{3}}C_{0}-\frac{1}{2}C_{2}-\sqrt{\frac{2}{3}}C_{3}\right|^{2}.
\end{eqnarray}
Since both the probability sum $P_{\mathrm S}$ and the probability $P_{\mathrm {WW}}(a,a)$ are expressed in terms of the same probability amplitudes in Eqs. (\ref{eq:PS1}) and (\ref{eq:paa3}), respectively, these two equations describe a fundamental relation between these two probabilities. As shown in Sec. \ref{sec:paradox}, a probability of $P_{\mathrm S}=0$ necessarily requires a probability of $P_{\mathrm {WW}}(a,a)=1/12$. This result can now be reproduced by considering Eq. (\ref{eq:PS1}), where $P_{\mathrm S}=0$ requires amplitudes of zero for $C_{1}$, $C_{2}$ and $C_{3}$, leaving $C_{0}=1$ as the only possible input state. Eq. (\ref{eq:paa3}) then confirms the probability of $P_{\mathrm {WW}}(a,a)=1/12$.

The relation between the probabilities $P_{\mathrm S}$ and $P_{\mathrm {WW}}(a,a)$ becomes more complicated when $P_{\mathrm S}$ is not zero. In that case, Eq. (\ref{eq:PS1}) does not have a unique solution and many different values for the coefficients $C_i$ $(i=0, 1, 2, 3)$ are possible. In general, $P_{\mathrm S}>0$ requires that $|C_{0}|<1$, with non-zero values for $C_{1}$, $C_{2}$ and $C_{3}$. This will result in a change of $P_{\mathrm {WW}}(a,a)$ due to the interferences between the eigenstates $|\phi_{0}\rangle$, $|\nu_{2}\rangle$ and $|\nu_{3}\rangle$ of the operator $\hat{\Pi}_{\mathrm S}$. It is worth noting that these interferences are describes by negative signs for the coefficients $C_{2}$ and $C_{3}$ corresponding to the eigenstates $|\nu_{2}\rangle$ and $|\nu_{3}\rangle$, indicating that the interferences with $|\phi_{0}\rangle$ are destructive whenever the coefficients are all real and positive. We can therefore conclude that a small increase of $P_{\mathrm S}$ can result in a much larger decrease of $P_{\mathrm {WW}}(a,a)$ as a result of destructive interferences between the probability amplitudes in Eq. (\ref{eq:paa3}).

\section{Statistical bounds for different measurement contexts}
\label{sec:statistcalbounds}

Eq. (\ref{eq:paa3}) shows how destructive interferences between$|\phi_{0}\rangle$ and a linear combination of $|\nu_{2}\rangle$ and $|\nu_{3}\rangle$ can reduce the probability $P_{\mathrm {WW}}(a,a)$ to values below $P_{\mathrm {WW}}(a,a)=1/12$. The magnitude of the effect depends on the probability amplitudes $C_{2}$ and $C_{3}$, requiring a non-zero contribution of $|\nu_2\rangle$ and $|\nu_3\rangle$ to the probability sum $P_{\mathrm S}$ shown in Eq. (\ref{eq:PS1}). We can now identify the quantitative relation between the possible reduction of $P_{\mathrm {WW}}(a,a)$ and the necessary probability sum $P_{\mathrm S}$ by finding the minimal value of $P_{\mathrm {WW}}(a,a)$ for a given value of $P_{\mathrm S}$. In particular, we can find out how high $P_{\mathrm S}$ must be to avoid a violation of the inequality in Eq. (\ref{eq:paa1}), providing a quantitative measure of the contradiction between non-contextual logic and the contextuality of the Hilbert space formalism for incompatible measurements.

As a first step, we observe that the state $|\nu_{1}\rangle$ is orthogonal to the state $|a,a\rangle$, which means that the state $|\nu_{1}\rangle$ is a joint eigenstate of the operators $|a,a\rangle\langle a,a|$ and $\hat{\Pi}_{\mathrm S}$ with eigenvalues of 0 and 1/2, respectively. In a superposition of $|\phi_{0}\rangle$ and $|\nu_{1}\rangle$, the probability amplitude $C_{1}$ reduces $P_{\mathrm {WW}}(a,a)$ by $(1/12)|C_{1}|^2$ while increasing the probability $P_{\mathrm S}$ by $(1/2)|C_{1}|^2$. Quantum interference effects will reduce $P_{\mathrm {WW}}(a,a)$ more rapidly for the same ``cost'' of increased $P_{\mathrm S}$. We therefore conclude that the states with the minimal value of $P_{\mathrm {WW}}(a,a)$ for a given value of $P_{\mathrm S}$ always have $C_{1}=0$. These optimal states can thus be found by distributing the probability $P_{\mathrm S}$ between the eigenstates $|\nu_{2}\rangle$ and $|\nu_{3}\rangle$. It is convenient to express this distribution by trigonometric functions of a parameter $\theta$, so that the probability amplitudes can be given by
\begin{eqnarray}
\label{eq:C2}
C_{2}&=&\sqrt{P_{\mathrm S}}\cos\theta \nonumber\\
C_{3}&=&\sqrt{\frac{2}{3}P_{\mathrm S}}\sin\theta.
\end{eqnarray}
Inserting Eq. (\ref{eq:C2}) into Eqs. (\ref{eq:PS1}) and (\ref{eq:paa3}), we can express the probability $P_{\mathrm {WW}}(a,a)$ as a function of the probability sum $P_{\mathrm S}$ and the parameter $\theta$ expressing the balance between the contributions of $|\nu_{2}\rangle$ and $|\nu_{3}\rangle$. Since Eq. (\ref{eq:paa3}) describes an interference effect between $C_{0}$ and a linear combination of $C_{2}$ and $C_{3}$, it is convenient to express the result in terms of two amplitudes associated with these components,
\begin{equation}
\label{eq:paa5}
P_{\mathrm {WW}}(a,a)=\left(A_{\mathrm 0}(\theta)-A_{\mathrm S}(\theta)\right)^{2}.
\end{equation}
Here, $A_{\mathrm 0}(\theta)$ represents the contribution of $C_{0}$. It depends on $\theta$ because the reduction of $C_{0}$ depends on the total magnitude of $C_{2}$ and $C_{3}$. Expressed as a function of $P_{\mathrm S}$ and $\theta$, the amplitude $A_{\mathrm 0}(\theta)$ is given by
\begin{equation}
\label{eq:AS1}
A_{\mathrm 0}(\theta)=\sqrt{\frac{1}{12}\left(1-P_{\mathrm S}\left(1-\frac{1}{3}(\sin\theta)^{2}\right)\right)}.
\end{equation}
A much stronger dependence on $P_{\mathrm S}$ and $\theta$ is obtained for the amplitude $A_{\mathrm S}(\theta)$ representing the contributions of $C_{2}$ and $C_{3}$,
\begin{equation}
\label{eq:AS2}
A_{\mathrm S}(\theta)=\frac{5}{6}\left(\frac{3}{5}\cos\theta+\frac{4}{5}\sin\theta\right)\sqrt{P_{\mathrm S}}.
\end{equation}
We can now consider the dependence of Eq. (\ref{eq:paa5}) on the parameter $\theta$ to find the minimal value of $P_{\mathrm {WW}}(a,a)$ for a given probability sum $P_{\mathrm S}$. Although it would be possible to minimize Eq. (\ref{eq:paa5}) directly, the different dependences on $\theta$ in $A_{\mathrm 0}(\theta)$ and in $A_{\mathrm S}(\theta)$ allow us to derive an initial bound by separately optimizing $A_{\mathrm 0}(\theta)$ and $A_{\mathrm S}(\theta)$. This procedure has the advantage that we can better understand the origin of the bound.

At low values of $P_{\mathrm S}$, $A_{\mathrm 0}(\theta)$ is greater than $A_{\mathrm S}(\theta)$.
Eq. (\ref{eq:paa5}) thus indicates that the minimal value of $P_{\mathrm {WW}}(a,a)$ is obtained when the amplitude $A_{\mathrm 0}(\theta)$ is minimal and the amplitude $A_{\mathrm S}(\theta)$ is maximal. This means that the bound of $P_{\mathrm {WW}}(a,a)$ can be obtained from the two bounds for $A_{\mathrm 0}(\theta)$ and for $A_{\mathrm S}(\theta)$ given by
\begin{eqnarray}
\label{eq:Alimit}
A_{\mathrm 0}(\theta) &\geq& \sqrt{\frac{1}{12}\left(1-P_{\mathrm S}\right)},
\nonumber \\
A_{\mathrm S}(\theta) &\leq& \frac{5}{6}\sqrt{P_{\mathrm S}}.
\end{eqnarray}
Note that the bound we are deriving is only valid when $A_{\mathrm 0}(\theta)$ is greater than $A_{\mathrm S}(\theta)$, since the bound for $P_{\mathrm {WW}}(a,a)$ drops to zero when $A_{\mathrm 0}(\theta)$ can be equal to $A_{\mathrm S}(\theta)$. Using the bounds in Eq. (\ref{eq:Alimit}), this occurs at $P_{\mathrm S}=3/28$, a numerical value of about $0.1071$. The corresponding lower bound of $P_{\mathrm {WW}}(a,a)$ is therefore
\begin{equation}
\label{eq:bound1}
P_{\mathrm {WW}}(a,a)\geq\left(\sqrt{\frac{1}{12}(1-P_{\mathrm S})}-\frac{5}{6}\sqrt{P_{\mathrm S}}\right)^{2} \hspace{0.5cm} \mbox{for} \hspace{0.5cm} P_{\mathrm S}\leq3/28.
\end{equation}
It is easy to see that this bound is tight for $P_{\mathrm S}=0$, where the value of $P_{\mathrm {WW}}(a,a)$ can only be 1/12. However, we expect the bound to become less tight as the value of $P_{\mathrm S}$ increases, since we used separate optimization procedures for $A_{\mathrm 0}(\theta)$ and $A_{\mathrm S}(\theta)$. To see how tight the bound is at the maximal value of $P_{\mathrm S}=3/28$, it is convenient to determine the actual value of $P_{\mathrm S}$ at which Eq. (\ref{eq:paa5}) achieves a probability of $P_{\mathrm {WW}}(a,a)=0$. The condition we need to satisfy is
\begin{equation}
\label{eq:A0}
A_{\mathrm 0}(\theta)=A_{\mathrm S}(\theta).
\end{equation}
This condition also depends on $\theta$, so we need to identify the value of the parameter $\theta$ at which the lowest value of $P_{\mathrm S}$ is obtained. The $\theta$-dependent relation is
\begin{equation}
\label{eq:PS2}
P_{\mathrm S}\left[P_{\mathrm {WW}}(a,a)=0\right] = \frac{1}{5-\cos(2\theta)+4\sin(2\theta)}.
\end{equation}
The lowest value of the probability sum $P_{\mathrm S}$ at which $P_{\mathrm {WW}}(a,a)=0$ is possible is found at $\cos(2\theta)=-1/\sqrt{17}$ and the lower bound of $P_{\mathrm S}$ for $P_{\mathrm {WW}}(a,a)=0$ is given by
\begin{equation}
\label{eq:PSbound}
P_{\mathrm S}\left[P_{\mathrm {WW}}(a,a)=0\right] \geq \frac{1}{5+\sqrt{17}}.
\end{equation}
This bound is very close to the limit of $P_{\mathrm S} \leq 3/28$ in Eq. (\ref{eq:bound1}), indicating that the bound is already rather tight. Numerically, the lowest value of $P_{\mathrm S}$ at which $P_{\mathrm {WW}}(a,a)$ can be zero is at $0.1096$, only $0.0025$ higher than the value at which the bound in Eq. (\ref{eq:bound1}) drops to zero.

Even though the bound is already rather tight, it may be interesting to consider the possibility of tightening it even further by making use of the precise bound in Eq. (\ref{eq:PSbound}). As mentioned above, the bound is obtained at $\cos(2\theta)=-1/\sqrt{17}$, corresponding to a value of $\sin\theta=0.7882$. By comparing this value with the value of $\sin\theta=0.8$ that maximizes $A_{\mathrm S}$ in Eq. (\ref{eq:AS2}), we can conclude that the optimal value of $\sin\theta$ drops continuously from its initial value of $\sin\theta=0.8$ to a final value of $\sin \theta=0.7882$ as the probability sum $P_{\mathrm S}$ increases from zero to its maximal value of $0.1096$. The optimal value of $\sin\theta$ is never close to the value of zero that maximizes the value of $A_{\mathrm 0}$ in Eq. (\ref{eq:A0}). We can therefore identify a tighter limit for $A_{\mathrm 0}$ by adding the condition that $\sin\theta>0.7882$. This tighter bound of $A_{\mathrm 0}$ can be approximated by
\begin{equation}
A_{\mathrm 0}(\theta) \geq \sqrt{\frac{1}{12}\left(1-\frac{4}{5}P_{\mathrm S}\right)}.
\end{equation}
Note that the value of $4/5$ is used in place of a more precise numerical value of $0.7929$ obtained from $\sin\theta>0.7882$ in order to present a more concise formula. The effect of the small difference on the tightness of the bound is negligibly small. The tighter bound is then given by
\begin{equation}
\label{eq:bound2}
P_{\mathrm {WW}}(a,a)\geq\left(\sqrt{\frac{1}{12}\left(1-\frac{4}{5}P_{\mathrm S}\right)}-\frac{5}{6}\sqrt{P_{\mathrm S}}\right)^{2} \hspace{0.5cm} \mbox{for} \hspace{0.5cm} P_{\mathrm S}\leq 0.109489,
\end{equation}
where the limit of $P_{\mathrm S}=0.109489$ is only slightly lower than the actual value at which the bound drops to zero. The exact value is found between $0.109489$ and $0.109490$, so the lower value has been chosen to ensure that the bound is valid at the maximally allowed value of $P_{\mathrm S}$.

This tighter bound is still a bit lower than the actual lower bound determined from a joint optimization of $A_{\mathrm 0}(\theta)$ and $A_{\mathrm S}(\theta)$ since the bound overestimates the value of $A_{\mathrm S}(\theta)$ by not taking into account the reduction of $A_{\mathrm S}(\theta)$ caused by the lower value of $\theta$ needed to optimize the contribution of $A_{\mathrm 0}(\theta)$. However, the tighter bound is much closer to the tight bound in Eq. (\ref{eq:PSbound}). The more precise numerical value of the tight bound is $P_{\mathrm S}(P_{\mathrm {WW}}(a.a)=0) \geq 0.109612$. The difference between the upper limit of $P_{\mathrm S}$ in the tighter bound of $P_{\mathrm {WW}}(a,a)$ given by Eq. (\ref{eq:bound2}) and the tight bound of $P_{\mathrm S}$ in Eq. (\ref{eq:PSbound}) is only $0.000123$, a reduction of a factor of 20 when compared to the bound in Eq. (\ref{eq:bound1}). We conclude that Eq. (\ref{eq:bound2}) is sufficiently close to the actual lower bound of $P_{\mathrm {WW}}(a,a)$ for most practical purposes.

\begin{figure}[ht]
\begin{picture}(500,200)
\put(50,0){\makebox(400,200){
\scalebox{0.40}[0.40]{
\includegraphics{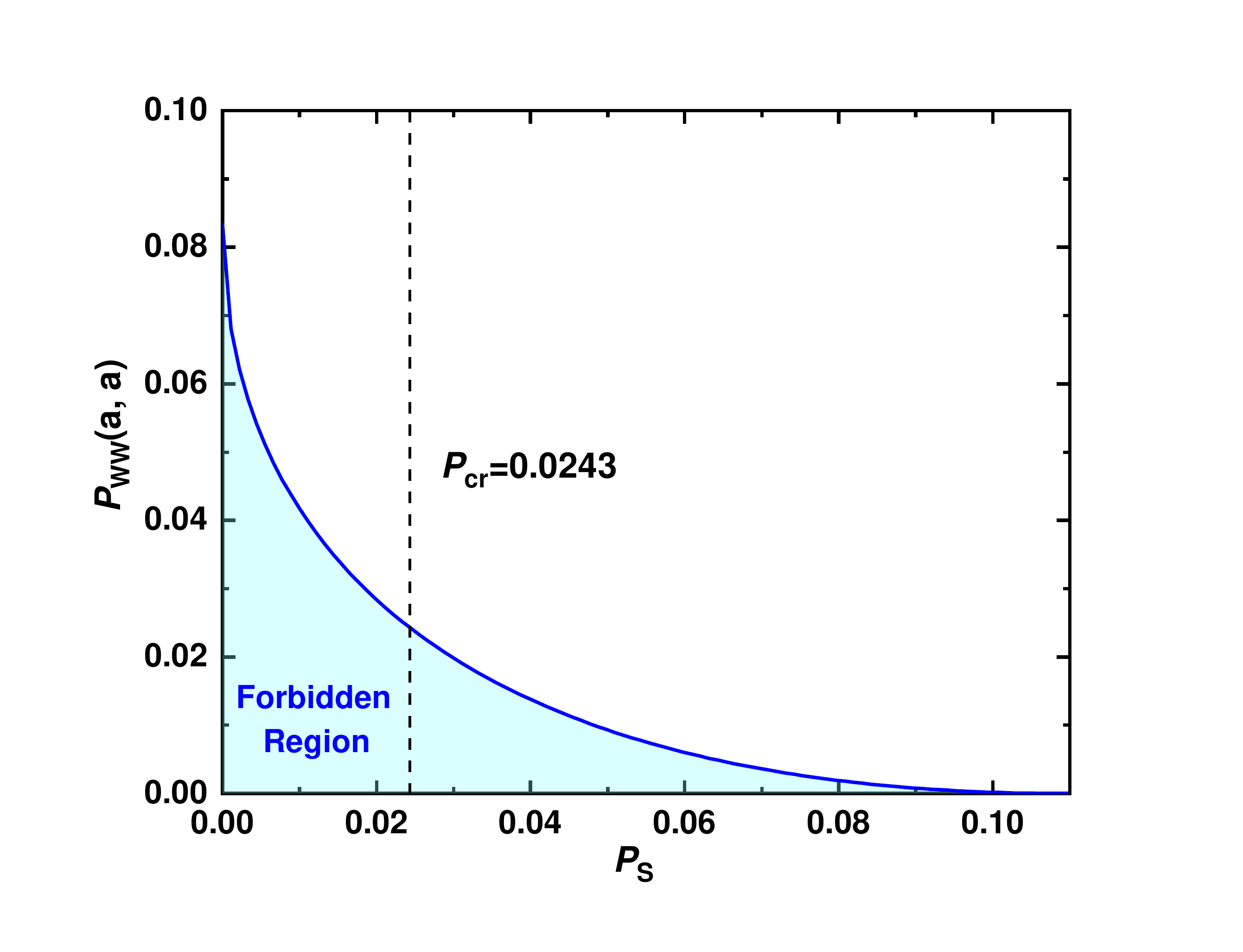}}}}
\end{picture}
\caption{\label{fig1}
Illustration of the lower bound of $P_{\mathrm {WW}}(a,a)$ given by Eq. (\ref{eq:bound2}). Quantum mechanics defines a ``forbidden region'' around the non-contextual expectation that $P_{\mathrm S}=0$ requires $P_{\mathrm {WW}}(a,a)=0$. The dotted line shows the value of the critical probability sum $P_{\mathrm {cr}}=0.0243$ below which quantum mechanics requires a violation of the inequality given in Eq. (\ref{eq:paa1}).
}
\end{figure}

The lower bound in Eq. (\ref{eq:bound2}) illustrates the fundamental relation between the probability sum $P_{\mathrm S}$ and the probability of the outcome $(a,a)$ described by the Hilbert space formalism. It directly contradicts the relation predicted by non-contextual logic and expressed by the inequality in Eq. (\ref{eq:paa1}). According to quantum mechanics, a probability sum of zero requires a probability of $P_{\mathrm {WW}}(a,a)=1/12$, and lower probabilities can only be achieved by destructive quantum interferences with the eigenstates $| \nu_{2}\rangle$ and $|\nu_{3} \rangle$ of the probability sum operator with eigenvalues of 1 and 3/2. Hilbert space thus defines a ``forbidden region'' around the result expected from the logical relation between non-contextual statements, where $P_{\mathrm S}=0$ requires $P_{\mathrm {WW}}(a,a)=0$. In fact, quantum mechanics even requires a minimal probability sum $P_{\mathrm S}\geq P_{\mathrm {cr}}$ in order to satisfy the inequality in Eq. (\ref{eq:paa1}). For all smaller probability sums, quantum mechanics necessarily violates this inequality. The critical probability sum $P_{\mathrm {cr}}$ at which $P_{\mathrm {WW}}(a,a)=P_{\mathrm S}$ is possible can be obtained from Eq. (\ref{eq:bound2}). Its numerical value is
\begin{equation}
P_{\mathrm {cr}}=0.0243.
\end{equation}
We can therefore conclude that quantum mechanics requires a violation of non-contextual logic whenever the sum probability $P_{\mathrm S}$ is below a value of 0.0243. A suppression of the sum of the probabilities of $(a,0)$, $(0,a)$ and $(1,1)$ below this value will necessarily result in a probability of the outcome $(a,a)$ that is higher than this sum.

The bound given by Eq. (\ref{eq:bound2}) and the critical probability sum $P_{\mathrm {cr}}$ are shown in Fig. \ref{fig1}. The minimal value of $P_{\mathrm {WW}}(a,a)$ required by low sum probabilities $P_{\mathrm S}$ drops quickly from its initial value of 1/12 at $P_{\mathrm S}=0$, achieving equality with $P_{\mathrm S}$ at the critical probability sum of $P_{\mathrm {cr}}=0.0243$. Below this critical probability sum, the relation between measurement outcomes described by state vectors in Hilbert space necessarily contradicts the logical relations between statements that do not depend on the measurement context. We can therefore conclude that quantum superpositions do not describe measurement independent realities even when each element of reality can be predicted with a precision approaching certainty. In fact, the bound derived above  shows that the expectation of a measurement independent reality necessarily fails in the limit of high precision, where the probability sum describing the rate of errors in the statements (1)-(3) is extremely low. The bound given by Eq. (\ref{eq:bound2}) and shown in Fig. \ref{fig1} successfully generalizes the observation in Sec. \ref{sec:paradox} that the relation between measurement outcomes represented by non-orthogonal quantum states is fundamentally different from the relation between measurement independent statements even when these statements can be verified with a success probability of one because they can be represented by orthogonality relations in Hilbert space.

\section{Conclusion}
\label{sec:conclusion}

We have investigated the non-classical relation between outcomes represented by non-orthogonal states in Hilbert space based on the contradiction between context-independent logic and quantum mechanics described by the consistency paradox. We observe that the contradiction is caused by the manner in which orthogonality relations in Hilbert space relate non-orthogonal states to each other. This difference also appears in the eigenstates and the eigenvalues of the sum of projection operators that described the sum of measurement probabilities obtained in separate measurements. We have analyzed the relation between the probability sum $P_{\mathrm S}$ associated with errors in the initial statement and the probability $P_{\mathrm {WW}}(a,a)$ associated with errors in the necessary conclusion by expressing the state $|a,a\rangle$ in terms of the eigenstates of the operator $\hat{\Pi}_{\mathrm S}$ representing the probability sum $P_{\mathrm S}$. The analysis shows that quantum interferences between the eigenstates are necessary to reduce the probability $P_{\mathrm {WW}}(a,a)$ below its value of 1/12 obtained for $P_{\mathrm S}=0$. Thus, the non-orthogonality of $|a,a\rangle$ and the eigenstates of $\hat{\Pi}_{\mathrm S}$ in Hilbert space not only results in a necessary violation of non-contextual logic, it also provides a precise quantitative bound assigning a minimal probability of $P_{\mathrm S}$ to values of $P_{\mathrm {WW}}(a,a)$ below $1/12$. A violation of the inequality required by non-contextual logic is necessarily observed whenever the probability sum $P_{\mathrm S}$ is below a critical value of 0.0243. We therefore conclude that the violation of non-contextual logic is an unavoidable consequence of the precise relations between outcomes represented by non-orthogonal states in Hilbert space.

The analysis presented above may have important implications for a wide range of quantum paradoxes. At the heart of our investigation is the rather unusual way in which probabilities are defined by quantum interference effects. Inequalities such as (\ref{eq:paa1}) can be violated because the statistics of quantum states do not correspond to joint probabilities of outcomes described by non-orthogonal states. It has been pointed out that the contextuality relations responsible for the violation of inequalities such as Eq. (\ref{eq:paa1}) correspond to the assignment of negative probabilities \cite{Spe08}, and these negative probabilities seem to be fundamentally related to the change of quantum coherences by unitary transformations \cite{Hof12,Hof15}. The correct conclusion seems to be that violations of inequalities are related to negative quasi-probabilities, establishing a link between paradoxes motivated by weak measurements \cite{Hof12,Aha91,Vai96,Aha13} and the inequality violations that indicate contextuality \cite{Lei05}. However, it is also clear that the negative values of quasi-probabilities are fundamentally linked to quantum interference effects. In the present analysis, we have provided a more direct link between the role of quantum interference effects in the definition of probabilities and the violation of an inequality associated with the assumption of non-contextual realism. We believe that this analysis can provide a useful starting point for a more systematic investigation of the role of quantum superpositions in defining the relation between the statistics observed in different measurement contexts.


\begin{thebibliography}{xyz00}


\bibitem{Bel64}
J. S. Bell, ``On the Einstein Podolsky Rosen paradox,'' Physics {\bf 1}, 195 (1964).
\bibitem{Koc68}
S. Kochen and E. Specker, ``The problem of hidden variables in quantum mechanics,'' Indiana Univ. Math. J. {\bf 17}, 59 (1968).


\bibitem{Cab08}
A. Cabello, ``Experimentally Testable State-Independent Quantum Contextuality,'' Phys. Rev. Lett. {\bf 101}, 210401 (2008).
\bibitem{Bad09}
P. Badziag, I. Bengtsson, A. Cabello, and I. Pitowsky, ``Universality of State-Independent Violation of Correlation Inequalities for Noncontextual Theories,'' Phys. Rev. Lett. {\bf 103}, 050401 (2009).
\bibitem{Kle12}
M. Kleinmann, C. Budroni, J.-A. Larsson, O. Guhne, and A. Cabello, ``Optimal Inequalities for State-Independent Contextuality,'' Phys. Rev. Lett. {\bf 109}, 250402 (2012).
\bibitem{Pan13}
A. K. Pan, M. Sumanth, and P. K. Panigrahi, ``Quantum violation of entropic noncontextual inequality in four dimensions,'' Phys. Rev. A {\bf 87}, 014104 (2013).
\bibitem{Ysu15}
H.-Y. Su, J.-L. Chen, and Y.-C. Liang, Demonstrating quantum contextuality of indistinguishable particles by a single family of noncontextuality inequalities,'' Sci. Rep. {\bf 5}, 11637 (2015).
\bibitem{Kun15}
R. Kunjwal and R. W. Spekkens, ``From the Kochen-Specker Theorem to Noncontextuality Inequalities Without Assuming Determinism,'' Phys. Rev. Lett. {\bf 115}, 110403 (2015).
\bibitem{Pxu16}
Z-P. Xu, D. Saha, H-Y. Su, M. Pawlowski, and J-L. Chen, ``Reformulating noncontextuality inequalities in an operational approach,'' Phys. Rev. A {\bf 94}, 062103 (2016).
\bibitem{Kri17}
A. Krishna, R. W. Spekkens, and E. Wolfe, ``Deriving robust non-contextuality inequalities from algebraic proofs of the Kochen-Specker theorem: The Peres-Mermin square,'' New J. Phys. {\bf 19}, 123031 (2017).
\bibitem{Kun18}
R. Kunjwal and R. W. Spekkens, ``From statistical proofs of the Kochen-Specker theorem to noise-robust noncontextuality inequalities,'' Phys. Rev. A {\bf 97}, 052110 (2018).
\bibitem{Sch18}
D. Schmid, R. W. Spekkens, and E. Wolfe, ``All the noncontextuality inequalities for arbitrary prepare-and-measure experiments with respect to any fixed set of operational equivalences,'' Phys. Rev. A {\bf 97}, 062103 (2018).
\bibitem{Lei20}
M. Leifer and C. Duarte, ``Noncontextuality inequalities from antidistinguishability,'' Phys. Rev. A {\bf 101}, 062113 (2020).


\bibitem{Bel66}
J. S. Bell, ``On the Problem of Hidden Variables in Quantum Mechanics,'' Rev. Mod. Phys. {\bf 38}, 447 (1966).
\bibitem{Gen05}
M. Genovese, ``Research on hidden variable theories: A review of recent progresses,'' Phys. Rep. {\bf 413}, 319 (2005).
\bibitem{Zel07}
F. De Zela, ``Single-qubit tests of Bell-like inequalities,'' Phys. Rev. A {\bf 76}, 042119 (2007).
\bibitem{Car18}
A. Carmi and E. Cohen, ``On the significance of the quantum mechanical covariance matrix,'' Entropy, {\bf 20}, 500 (2018).
\bibitem{Tem19}
T. Temistocles, R. Rabelo, and M. T. Cunha, ``Measurement compatibility in Bell nonlocality tests,'' Phys. Rev. A {\bf 99}, 042120 (2019).


\bibitem{Fra18}
D. Frauchiger and R. Renner, ``Quantum theory cannot consistently describe the use of itself,'' Nat. Commun. {\bf 9}, 3711 (2018).
\bibitem{Har92}
L. Hardy, ``Quantum mechanics, local realistic theories, and Lorentz-invariant realistic theories,'' Phys. Rev. Lett. {\bf 68}, 2981-2984 (1992).
\bibitem{Har93}
L. Hardy, ``Nonlocality for two particles without inequalities for almost all entangled states,'' Phys. Rev. Lett. {\bf 71}, 1665-1668 (1993).
\bibitem{Cab13}
A. Cabello, P. Badziag, M. T. Cunha, and M. Bourennane, ``Simple Hardy-Like Proof of Quantum Contextuality", Phys. Rev. Lett. {\bf 111}, 180404 (2013).
\bibitem{Mar14}
B. Marques, J. Ahrens, M. Nawareg, A. Cabello, and M. Bourennane, ``Experimental Observation of Hardy-Like Quantum Contextuality", Phys. Rev. Lett. {\bf 113}, 250403 (2014).
\bibitem{Mer94}
N. D. Mermin, ``Quantum mysteries refined,'' Am. J. Phys. {\bf 62}, 880-887 (1994).
\bibitem{Pro19}
M. Proietti, A. Pickston, F. Graffitti, P. Barrow, D, Kundys, C. Branciard, M. Ringbauer and A. Fedrizzi, ``Experimental test of local observer independence,'' Sci. Adv. {\bf 5}, eaaw9832 (2019).


\bibitem{Spe08}
R. W. Spekkens, ``Negativity and Contextuality are Equivalent Notions of Nonclassicality,'' Phys. Rev. Lett. {\bf 101}, 020401 (2008).
\bibitem{Hof12}
H. F. Hofmann, ``Complex joint probabilities as expressions of reversible transformations in quantum mechanics,'' New J. Phys. {\bf 14}, 043031 (2012).
\bibitem{Hof15}
H. F. Hofmann, ``Quantum paradoxes originating from the nonclassical statistics of physical properties related to each other by half-periodic transformations,'' Phys. Rev. A {\bf 91}, 062123 (2015).
\bibitem{Aha91}
Y. Aharonov, L. Vaidman, ``Complete description of a quantum system at a given time,'' J. Phys. A: Math. Gen. {\bf 24}, 2315 (1991).
\bibitem{Vai96}
L. Vaidman, ``Weak measurement elements of reality,'' Found. Phys. {\bf 26}, 895 (1996).
\bibitem{Aha13}
Y. Aharonov, S. Popescu, D. Rohrlich, and P. Skrzypczyk, ``Quantum Cheshire Cats,'' New J. Phys. {\bf 15}, 113015 (2013).
\bibitem{Lei05}
M. S. Leifer and R. W. Spekkens, ``Pre- and post-selection paradoxes and contextuality in quantum mechanics,'' Phys. Rev. Lett. {\bf 95}, 200405 (2005).

\end{thebibliography}
\end{document}